\documentclass{kapproc} 
\let\footnote\savefootnote
\let\footnotetext\savefootnotetext 
\let\footnoterule\savefootnoterule 
\kluwerbib
\def\hkpc{$h^{-1}{\ }{\rm kpc}$}
\def\hMpc{$h^{-1}{\ }{\rm Mpc}$}
\def\nbody{$N$-body}
\newcommand{\astroph}[1]{\mbox{\texttt{astro-ph/#1}}}

\RequirePackage{graphicx}%
\RequirePackage{epsf}%

\begin{document}

\articletitle{Cosmology on a Mesh}

\author{Stuart P.~D. Gill, Alexander Knebe, Brad K Gibson, Chris Flynn}
\affil{Centre for Astrophysics \& Supercomputing, Swinburne University,
Australia}
\email{sgill,aknebe,bgibson,cflynn@astro.swin.edu.au}
\author{Rodrigo A. Ibata}
\affil{Strasbourg Observatory, France}
\email{ibata@newb6.u-strasbg.fr}
\author{Geraint F. Lewis}
\affil{School of Physics, University of Sydney, Australia}
\email{gfl@physics.usyd.edu.au}

\chaptitlerunninghead{Cosmology on a Mesh}

\begin{abstract}
An adaptive multi grid approach to simulating the formation of structure
from collisionless dark matter is described. {\tt MLAPM} (Multi-Level
Adaptive Particle Mesh) is one of the most efficient serial codes
available on the cosmological ``market'' today.  As part of Swinburne
University's role in the development of the Square Kilometer Array, we are
implementing hydrodynamics, feedback, and radiative transfer within the
{\tt MLAPM} adaptive mesh, in order to simulate baryonic processes relevant to
the interstellar and intergalactic media at high redshift.  We will
outline our progress to date in applying the existing {\tt MLAPM} to a study of
the decay of satellite galaxies within massive host potentials.
\end{abstract}

{\tt MLAPM} (Multi-Level Adaptive Particle Mesh) is a publicly available 
C-code\footnote{\tt http://astronomy.swin.edu.au/MLAPM/}
for evolving a set of $N$-particles under their mutual gravity within a
cosmological framework. The code solves Poisson's equation on a
hierarchy of nested grids; the entire computational volume is covered by
one cubic domain grid, while refined regions are of arbitrary shape and
adjusted to the actual density field at each major time-step in order to follow
the real distribution of particles at all times. An example of {\tt MLAPM} in
action is shown in Figure~1. The left panel shows all particles in a slice
of thickness 3\hMpc\ through the simulation box. The right panel
indicates the adaptive grids used with that particle distribution. In
addition to this spatial refinement, an additional adaptive
time-stepping is implemented in the latest version of {\tt MLAPM}.  
The time stepping is restricted so that we ensure that particles are 
advanced at least a pre-specified fraction of the cell in which it resides, but
never more than half the cell spacing.
 
\begin{figure}[h]
\vspace{48mm}
\includegraphics{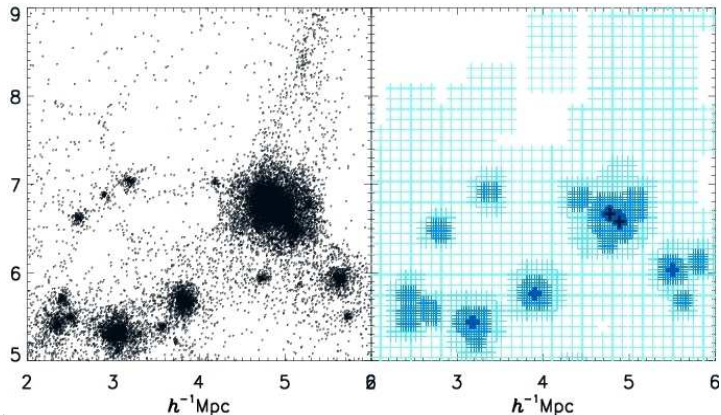}
\caption{\footnotesize {\label{fig:refine} Particle distribution (left)
vs. {\tt MLAPM} adaptive grids (right)}}
\end{figure}


{\tt MLAPM} has proven to be one of the
fastest single-CPU \nbody-codes on the market today (Knebe et~al. 2001). 
We have begun an ambitious program of cosmological and galactic dynamical
simulations using {\tt MLAPM} - preliminary results on the latter are presented here.

The signatures of hierarchical galaxy formation are evident in the
observed substructure seen in various phase-space projections of the Galactic
halo.  The clearest such signature is that of the spectacular stream of
stars associated with the currently disrupting Sagittarius Dwarf Galaxy.
Secondary streams have also been observed locally (Helmi et~al. 1999)
and in the halo of M31 (Ibata et~al. 2001). Ibata et~al have shown that
such streams are extremely useful tools for constraining the shape of a
halo's gravitational potential well.  In the case of the Milky
Way, Ibata et~al. concluded that our halo was necessarily very close to spherical (under
the assumption of a static axisymmetric potential). We have adopted {\tt MLAPM}
to extend this analysis, but now are using live potentials.

Four low resolution ($128^{3}$ particles; 64\hMpc\ box size) 
simulations were run initially, and ten halos 
selected sampling a range of triaxialities.  These
halos were then re-simulated at higher resolution ($512^{3}$ particles).
The effective mass per particle was 10$^{6}$~M$_{\odot}$, with a force
resolution of 1\hkpc\ in the dense regions.  An adaptation of the
Bound Density Maxima (BDM, Klypin \& Holtzman 1997) algorithm was used to
identify and trace substructure evolution in these high resolution
simulations.

Our preliminary results show that triaxiliality is a fleeting measure of the
Galactic potential - the live potential and active substructure mitigates
(somewhat) the usefulness of this measure.  One simulated stream is
highlighted here in Figure~2.  A detailed analysis of the phase-space dissolution of
these structures is currently underway.

\begin{figure}[h]
\vspace{60mm}
\includegraphics{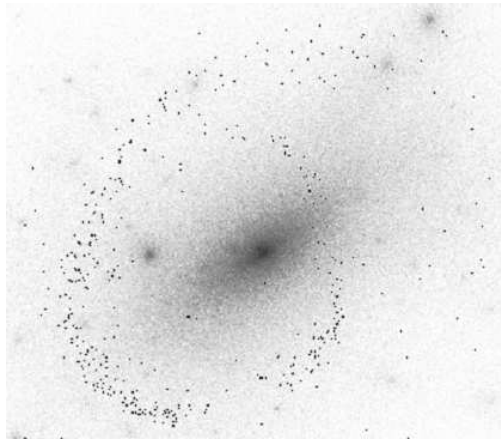}
\caption{\footnotesize {\label{fig:halo} Disrupted satellite in halo potential}}
\end{figure}

We have begun implementing hydrodynamics within {\tt MLAPM}, and aim to have a
publicly available version of {\tt Hydro-MLAPM} in 2004-2005.
The existing grid structure will be used as a
base for this implementation, as the grid provides a natural structure on
which to solve the relevant equations.



\end{document}